# Improving time use measurement with personal big data collection - the experience of the European Big Data Hackathon 2019

*Mattia Zeni1, Ivano Bison2, Britta Gauckler3, Fernando Reis4, and Fausto Giunchiglia5*

This article assesses the experience with i-Log at the European Big Data Hackathon 2019, a satellite event of the New Techniques and Technologies for Statistics (NTTS) conference, organised by Eurostat. i-Log is a system that allows to capture personal big data from smartphones' internal sensors to be used for time use measurement. It allows the collection of heterogeneous types of data, enabling new possibilities for sociological urban field studies. Sensor data such as those related to the location or the movements of the user can be used to investigate and have insights on the time diaries' answers and assess their overall quality. The key idea is that the users' answers are used to train machine-learning algorithms, allowing the system to learn from the user's habits and to generate new time diaries' answers. In turn, these new labels can be used to assess the quality of existing ones, or to fill the gaps when the user does not provide an answer. The aim of this paper is to introduce the pilot study, the i-Log system and the methodological evidence that arose during the survey.

*Key words:* time use survey; big data; ubiquitous computing; smartphones; smart surveys.

*Acknowledgments*: This research has received funding from the European Union's Horizon 2020 FET Proactive project "WeNet – The Internet of us", grant agreement No 823783.

## 1. Introduction

In October 2018 the official statistics offices of the European Statistical System (ESS) agreed on the "Bucharest Memorandum on Official Statistics in a datafied society (Trusted

1 Department of Information Engineering and Computer Science, University of Trento, via Sommarive 9, 38123 Povo, Trento, Italy. Email: mattia.zeni.1@unitn.it
2 Department of Sociology and Social Research, University of Trento, via Verdi 26, 38123, Trento, Italy. Email: ivano.bison@unitn.it
3 European Commission – DG EUROSTAT, 5 rue Alphonse Weicker, Luxembourg. Email: britta.gauckler@ec.europa.eu
4 European Commission – DG EUROSTAT, 5 rue Alphonse Weicker, Luxembourg. Email: fernando.reis@ec.europa.eu
5 Department of Information Engineering and Computer Science, University of Trento, via Sommarive 9, 38123 Povo, Trento, Italy. Email: fausto.giunchiglia@unitn.it

Smart Statistics)" (European Statistical System Committee 2018). This memorandum was a major evolution with respect to the Scheveningen Memorandum (European Statistical System Committee 2013). The ESS committed itself to a set of actions towards the implementation of changes in the way official statistics are produced, with the goal of continuing to fulfil its role in a society where data, smart technologies and artificial intelligence are a reality. As part of this effort, big data is an important potential additional source for the production of official statistics.

One of the pillars of smart statistics are trusted smart surveys. Smart surveys are enabled by personal devices, equipped with sensors and mobile applications that combine two data collection modes: one based on active inputs from the subjects (e.g., responses to queries, shared images), and another based on the data collected passively by the device sensors (e.g. accelerometer, GPS). Trusted smart surveys augment smart surveys with a set of technologies which together increase privacy preservation and data security, enhancing their degree of trustworthiness and therefore acceptance by the citizens (Eurostat 2019b).

The Harmonised European Time Use Survey (HETUS) is one of the European official statistics tools that could take advantage of an implementation as a trusted smart survey. Given the novelty of the use in official statistics of data from sensors available in smartphones, it was important to kick-start with an exploration of the possibilities of use in the context of time use measurement. The tool chosen for this exploration was a hackathon, where a significant number of competing teams attempted, in a short period of time, to find solutions to a statistical challenge.

This paper reports on the experience acquired during the European Big Data Hackathon 2019, as the basis for further future development towards more evolved trusted smart surveys. The remainder of this paper is organized as follows: Section 2 presents the state of the art, Section 3 presents the i-Log system (Zeni et al. 2014), that has been used to perform the pilot study. Section 4 details the data collection and preparation, Section 5 describes how i-Log pilot studies are organized and carried out. Section 6 lays out the specific use case of the European Big Data Hackathon 2019. Section 7 draws the lessons learned from the pilot study, and finally Section 8 presents the conclusions and summarises the main findings of this paper.

2. **Time Use Surveys during the Internet Era**

The aim of Time Use Surveys (TUS) is the measurement of time use by individuals and households. In more detail, TUS measure the frequency and duration of human activities, offering a detailed view of the social behaviour of members of society. Finally, they allow us to understand how certain variables influence the use of time (Dumazedier 1975) by households and household members.

TUS are more than just frequencies of individual- and group time use, since they can be used as "*a unique tool for exploring a wide range of policy concerns including social* *change; division of labour; allocation of time for household work; the estimation of the value of household production; transportation; leisure and recreation; pension plans; and health-care programmes, among others*" (United Nations 2010). They can help answering different questions of social and economic relevance, like revealing the living conditions of a society and identifying societal changes, as well as allowing us to measure living standards within a population and between countries. Furthermore, they provide information about the demand by citizens for public and private services that are of high relevance in the decision-making process and social planning. They also allow to upgrade economic accounts, improve labour force analysis, support the evaluation of social change, the study of gender issues, the progress on the improvement of quality of life, and a systematic analysis of leisure time (Robinson 1999).

TUS collect two types of information. The first is diachronic, i.e. underlying activity sequences in time episodes (e.g. of ten minutes) over a period of one day up to a week. Such type of data is usually collected by a self-completed time-diary that allows registering, at fixed time intervals, the sequence of an individual's activities. For each main activity in each interval, additional information is usually recorded, like a secondary activity and information about "where" and "with whom" this activity was done. The second type of information collected within a TUS is synchronic, that includes a paper-based or computer-assisted set of personal interviews (CAPI) about socio-economic individual and household background variables and often regarding different aspects of the household and people's wellbeing. Usually, specific information is included in the questionnaire about less frequent activities for a period longer than a day and/or item-specific questions like a seven-day work schedule. The seven-day work schedule proposed by HETUS (Eurostat 2009; Merz 2009), was removed from the guidelines of Eurostat 2019 because it is used neither by a large majority of the participating countries, nor by Eurostat. However, as stated in the guideline "*The weekly schedule of working time can be re-introduced into HETUS when … new technical solutions will be available for the survey*" (Eurostat 2019).

Recently, two main aspects posed new challenges to TUS (Juster and Stafford 1991). Firstly, changes in people's living conditions and resulting use of time require adaptions to TUS. In this regard, the balance between in-home and out-of-home time remained mainly unchanged (Gershuny and Sullivan 2019) in the last decades. However, the time people spend online increased considerably, while the offline time spent on social activities, activities with other people, reading books and newspapers and offline hobbies declined (Vilhelmson et al. 2018; Juster et al. 2014).Secondly, the increasing interest of the academic research community poses new requirements to TUS (Juster and Stafford 1991).

The three main challenges TUS face today are (1) the ability to capture the complexity of social life completely, (2) the granularity of the information, and (3) the cost to run such a study, both in terms of money and time. In the last three decades, there has been an increasing interest by the research community in TUS. There is new interest in investigating the sequence of the activities and the time of the day at which activities occur. Simultaneous activities can be properly investigated, and if multiple diaries are collected within a single household, researchers can use them to investigate patterns of co-presence, interdependence and cooperation ([Gershuny 2015](#)). Research results obtained through TUS pertain to three main thematic areas: (1) debates on the leisure civilisation and the end of work; (2) work and life rhythms; and, (3) intra-familial synchronising of social time ([Chenu and Lesnard 2006](#); [Bison and Scalcon 2018](#).

At the same time, computers and modern technologies have completely changed the types of activities that should be recorded. Related to the growing availability of new technologies, the question whether it is more important to measure the time spent on the computer or to capture the activities done at the computer (e-mailing, researching, reading, chatting, etc.) or on digital media has been raised ([Kramarczyk 2015](#)). For example, on average in 2018, U.S. adults spent over 11 hours a day connected to linear and digital media, performing different activities like watching, reading, listening to or simply interacting with media, according to the Q1 2018 study by market-research group Nielsen ([The Nielsen Company 2018](#)). This increase in the time spent on new technologies is not only due to the younger generations. For instance, in Q1 2018, younger adults (18-34 years old) spent less than nine hours a day, as compared to older adults (50+) who spend over 12 hours a day with content available across platforms, with a maximum for adult 50-64 of 12:50 hours a day. On the other side, young adults 18-34 spend 57% of their time-consuming media on digital devices (App/Web on a Tablet/Smartphone, Internet on a Computer and TV-Connected Devices.

Internet is changing the individual and societal perception of time and space ([Castells 2000](#); [Kramarczyk and Osowiecka 2014](#)). In this perspective: (a) geographic distances are losing their importance and abolishing the distinction between leisure and work, making the division between family, friends and work transparent. (b) The amount of time dedicated to each activity is reduced due to the time compression ([Barney 2004](#)), i.e. the ability to perform multiple tasks at the same time. For example, while traveling by train, it is now possible to connect to the internet, make a commercial transaction, send an e-mail, eat a sandwich, watch a movie, meet on/offline friends. Paradoxically, on the one hand, new technologies lead to saving time and, at the same time, the increasing importance in our everyday life of such activities make them time-consuming ([Kramarczyk and Osowiecka 2014](#)). Nowadays, spatial mobility requires new and more in-depth information. It is not enough anymore to capture the travel event and the reason for that. It is also crucial to understand peoples' travel behaviours. It is not enough to know only the origin and destination of the trip, but it is also important to know, e.g., the route and the time taken. Moreover, in a multiplicity of tasks carried out at the same time, it is increasingly important to have more detailed information. For example, computers are a means of carrying out an activity (e.g. office work) but they can also replace an activity, while the activity that has been replaced is also essential (e.g. work during a train journey). Therefore, it is important to collect both information so that the researcher has full flexibility depending on the research question.

Finally, an important challenge concerns the frequencies of the observations and the time at which they are carried out. In fact, increasingly often, there is a higher demand for a faster provision of data that are of high relevance in the decision-making process and social-economic planning or the measuring of well-being ([United Nations 2010](#)). However, on the other hand, due to the high cost and the complexity, especially for work required to process the collected data, e.g., the correct coding of open answers by dedicated coders ([Hellgren 2014](#)), most of TUS taketime intervals of around ten years. Notice, how typical intra-personal issues such as *social well-being, work-life balance, use of information and communication technologies, mobility and travel, physical activity, social environment, geographical context, regularity and frequency of individual activities* cannot be studied if they are observed only for one day, but need to be observed for more, consecutive days, for a typical period and at household level.

In a nutshell, the future challenge is to introduce new methods and technologies to conduct a TUS that allows, for instance, *new ways of sampling time use*, *to record information that combines automatic and continuous data collection*, *with/without human intervention*, *that is more accurate in data collection by leveraging the new opportunities that technology offers*. Through their introduction, more context sensitive data could be collected, the burden of filling out a traditional diary could be reduced, and overall expenses lowered.

To answer most of these challenges, one opportunity comes from new human mobile technologies such as smartphones and the applications they run, or any other type of wearable device (e.g., smartwatches). The smartphone has become an integral part of the life of large parts of the population, both in economically advanced countries and in developing countries. Over time, more and more people are using smartphones all the time and they are using them, for instance, to send text messages, to be active on social media, to check the news, to find places on a map and (even) to call other people. The Mobile Economy report Europe 2018 ([GSM Association 2018](#)) forecasts that by 2025 the penetration rate of connected devices on the European population will be 88%, with individual subscriptions (SIM cards) at 128% and the smartphone adoption, as a percentage of the total connections, will be 83%.

These technologies are a valuable alternative to traditional paper diary instruments used for surveying and they allow time use research to be carried out in a completely different way ([Fernee and Sonck 2014](#)). Smartphones not only allow respondents to report their activities at a finer grade pace per day and over multiple days, but also enable the collection of complementary information, such as the person's mood or how people feel at random moments during the day (e.g. experience sampling), what short-term activities they do throughout the day, etc. Moreover, smartphones are a perfect tool for collecting multiple types of 'passive' data, such as geospatial or inertial sensor streams (from GPS and inertial sensors), and for collecting the interactions or communications with others (by monitoring social media apps, calling, voice, text, SMS, email, video-chat but also using Bluetooth-enabled measurements). Finally, they allow us to collect data about how people use smartphones (by use of specialized applications supporting, e.g., visual data collection, audio recording, scanning, taking pictures, listening to music, visiting social network sites). Even more interesting is how these types of data can be combined with data collected with other modalities (e.g., personal and household questionnaire by Computer Assisted Web Interviews (CAWI), smartphone beeper/ notifications that collect information at regular points in time including the time diary information ([Robinson 2002](#)), and continuous data from sensors). The result is a much more comprehensive overview of the respondent's time-use, behaviour and well-being ([Fernee et al. 2013](#); [Fernee and Sonck 2014](#)). In this way, smartphones are not simply a replacement for the traditional paper and pencil time use diaries, but a 'multifunctional tool' that allows us to combine the traditional methods with new data sources which would not be possible without smartphones ([Link et al. 2014](#)).

Of course, smartphone survey research is rather new, and there are only a few early examples of applications of their use in time use surveys ([Fernee et al. 2013](#); [Giunchiglia et al. 2017](#); [Giunchiglia et al. 2018](#)). Furthermore, how pointed out by [Link (2018)](#), many and new methodological and technical problem arise, from the sampling mode and the penetration rate, to the ethics and privacy matters, to the usability, connectivity, design and layout of the app, but also the battery life and the operating systems. Briefly, the "who, what, when, where, and why" of smartphone usage can vary dramatically.

Conversely, only now we can start to imagine the potential and the opportunity that this new way of data collection for the scientific community and other stakeholders to increase the knowledge about human behaviour and social rhythms. Paraphrasing and reversing the suggestion provided by [Groves (2011)](#), with a smartphone, we now collect "Organic Data" supplemented by "Designed Data": a fruitful combination of behavioural data from sensors and self-report data from human respondent. In the last two decades, the smartphone has opened the door to a new generation of measurement tools for those who study public opinion, attitudes and behaviours as well as other sociological phenomena ([Link et al. 2014](#)). They enable researchers to collect information that was previously unobservable or difficult to measure, expanding the realm of empirical investigation ([Sugie 2018](#)). With the new functionality of a smartphone, we can capture information on people's attitudes, surroundings, interactions, and behaviours to gain a rich gratitude for the different lifestyles and personalities that characterize a particular population. In this scenario, "*the use of multimode data-collection apps is not simply the next stage in evolution of CAI, but rather a species unto itself, with elements of CAI interacting with a new set of user expectations.*" ([Link et al. 2014](#)).

3. **The i-Log System**

i-Log ([Zeni et al. 2014](#)) is a system used to carry out data collection campaigns with the ultimate goal of studying different aspects of human behaviour related to the use of time. The system is composed of (1) a backend infrastructure deployed in the cloud, designed to handle huge number of users and workloads. It is responsible for collecting, processing, storing and making the collected data available for further analysis. The second component is (2) a mobile application that runs on the users' personal mobile devices.

The mobile application has been created for Android mobile devices (iOS version is currently under development) and allows to continuously collect data about the user. More in detail, two different types of data are generated: streams of personal big data from the smartphone's internal sensors and time diaries in the form of answers to specific questions. This duality of data types makes i-Log unique with respect to other tools currently available ([Runyan et al. 2013](); [Wang et al. 2014](); [Hatuka and Toch 2017]()) and allows to address new challenges that concern the sociological and urban fields, in three ways. First of all, it allows to investigate the real world through data recorded by phone sensors, e.g., geolocation. Secondly, it allows to improve existing time diaries ([Sorokin and Berger 1939](); [Zeni 2017]()), especially for structured ones ([Hellgren 2014]()). Generally, the problem of time diaries is that they are expensive and time consuming, both for the participants to fill them and for the researcher, to analyse the data. i-Log can help in this regard because of its ability to produce highly standardized and comparable survey results. Each answer to the survey is mapped automatically to a hierarchy of concepts collected in an ontology based on WordNet. In this way, even if the survey is provided in different languages, the output is always composed by a set of standardized concepts that do not need manual processing. Finally, the third advantage enabled by i-Log is that it can help the respondent in providing the answer, i.e., by reducing her cognitive load. In fact, it can compensate for gaps in the data due to the subject's attention and memory deficits that appear in traditional measurement tools. This is solved through the combination of data that i-Log collects, which in principle allow the training of machine learning models using the time-diaries answers as annotation labels. As a result, the trained models can be used to generate labels when these are not available, filling the gaps in the data.

i-Log is operational for a data collection experiment in an uncontrolled field environment, outside laboratory settings and with unexperienced users. Its main characteristics in this regard are:
- **Optimized battery usage:** today's smartphones are powerful devices with hardware characteristics comparable to high-end personal computers. Additionally, they are empowered by an operating system that is designed precisely to run applications that allow it to be used in almost any circumstance. However, this did not come without costs: the energy consumption increased significantly. The fact that in the past decade there is no major breakthrough in the battery technology highlights that the battery is currently the main limitation of today's smartphones. The main solution that smartphone and application producers found is to limit the execution time of the applications on the devices as much as possible, every year with a more aggressive solution. Therefore, creating an application that runs continuously and in an efficient way is particularly challenging. i-Log runs continuously in the background to collect sensor data from the device, without creating a major impact on the battery life.

- **Generation of Truthful data:** in order to collect truthful data from the users, we decided to install i-Log on their personal devices instead of providing dedicated ones. This choice has different advantages, starting from cost reduction, to speeding up the set-up time of an experiment among others, but also multiple challenges. A big challenge was to reduce the impact of the usage of the tool on the users' smartphones as much as possible. In fact, a user who realises her/his device slowed down, or impacted by our application, would have quitted the experiment immediately, or have altered the normal use and consequently altering the collected data. For this reason, we put a lot of effort into the simplification and improvement of i-Log performances. We removed the user interface typical of most of the applications on the market completely, and instead used an approach based on *notifications*. In fact, the user was still informed about the data collection process, but we decided to do so through a non-invasive notification present in the notification area of the device (Figure 1). The user can understand that i-Log is running and collecting data and perform some basic actions like stopping the data collection (another very important aspect related to privacy and ethics, the user should always be in control), or opening the settings and questions menu. The only situation in which the user is required to interact with an UI element in i-Log is when time diaries are filled in. These questions are downloaded from a remote server at specific time intervals as per the experiment characteristics and have specific formats.

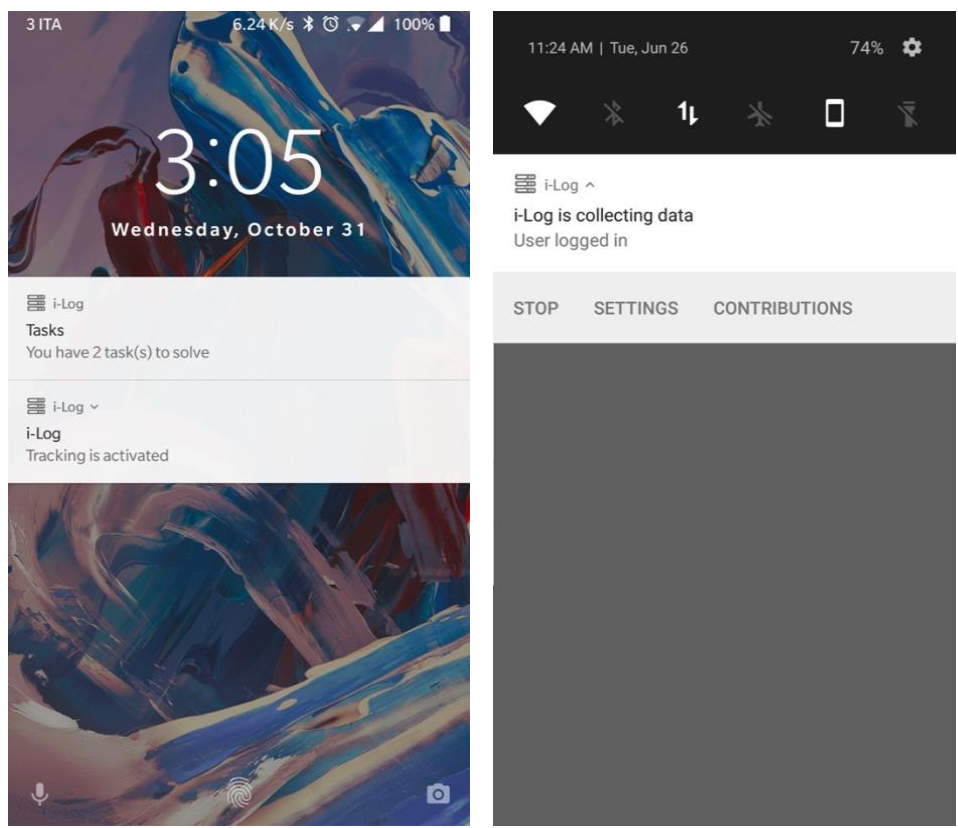

**Figure 1**. i-Log notification system. The first notification is always on and is used to inform the user about the data collection ("Tracking is activated"). The second one instead is present when time-diaries are available to be filled ("You have 2 task(s) to solve").

At the moment, i-Log allows to reply to different combinations of types of questions/answers:

- Text question, multiple choice answer (Figure 2a)
- Text question, single choice answer
- Text question, open text answer
- Text question, map (component)
- Map (component) question, multiple-choice answer, i.e., what were you doing in the location selected on the map below? (Figure 2b)
- Image question, multiple-choice answer, i.e., what do you see in this picture?

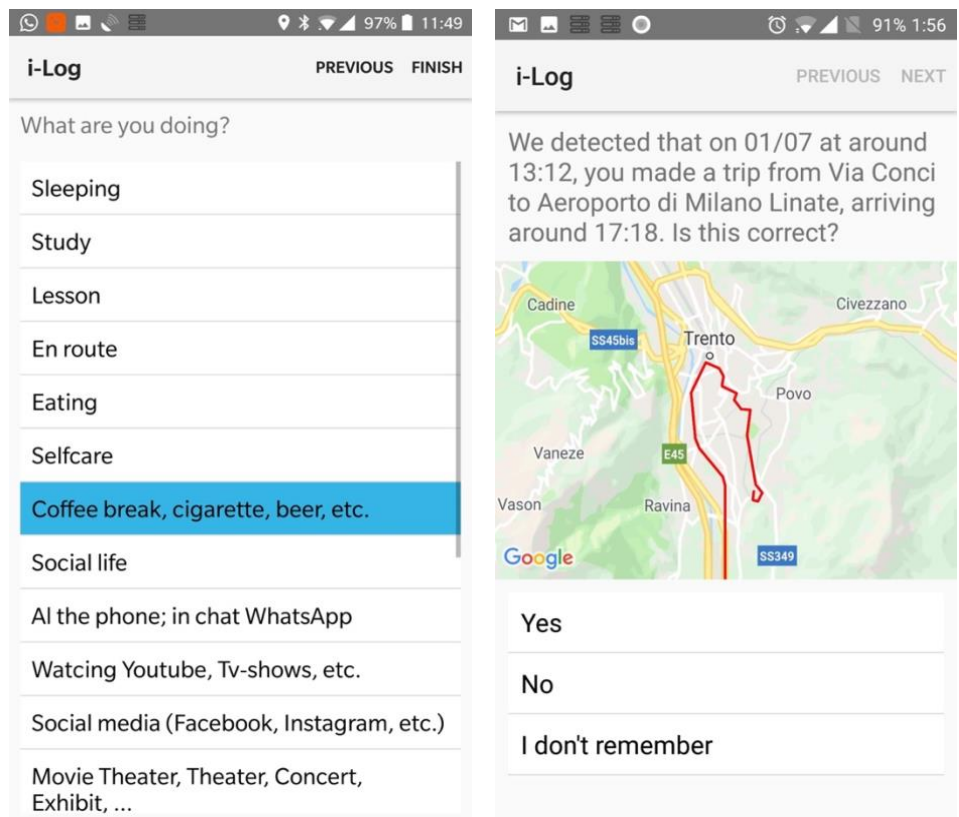

**Figure 2**. i-Log user interface about text questions (left) and map questions (right).

- **Low resource usage:** The current version of i-Log has been designed to run on Android, but an iOS version is currently in development. Android is the most adopted operating system worldwide and runs on thousands of different devices. To accommodate the requirements of most of them we had to reduce the resources (CPU and RAM) i-Log uses as much as possible by optimizing its code and delegating part of the intensive tasks to the backend. This as an obvious advantage also in terms of energy use.
- **Ability to work Offline:** i-Log has been designed to work offline. It can perform most of its tasks even if the phone is temporarily not connected to the internet. This is indeed what happens in reality, a smartphone is always connected except for some specific situations, i.e., the user is in a tunnel, in the basement, in the metro. The data are collected locally and temporarily stored in a secure location in form of log compressed files. Periodically, these files are opportunistically sent over the network to the backend system that processes them.

## 4. Data collection and data preparation

Each smartphone is different: different brand, model, hardware components and different software versions. For such reasons, available sensors differ and have different characteristics (collection frequency, accuracy, reliability, etc.). This makes data collection on a smartphone a challenging task.

In i-Log, we can distinguish between two types of sensors, *hardware* and *software*. The previous refers to those physically embedded in the device, like the accelerometer, the gyroscope and the GPS, among others. The latter instead refers to software components that generate measurable features, such as an event when a new notification pops-up, or when the device connects to a Wi-Fi network. In i-Log, the sensor data collection process can be configured based on the needs of each pilot study. What can be configured are the sensors to collect data from and at which frequency to collect data from them, e.g., it can be decided to collect data from the accelerometer at a pace of 60 values per second, while not collecting from the gyroscope at all. The following table shows the complete list of sensors available at the time of writing this paper, together with their default collection frequency:

| Sensor | Frequency | Sensor | Frequency | Sensor | Frequency |
|---|---|---|---|---|---|
| Acceleration | 20Hz | Screen Status | On change | Proximity | On change |
| Linear Acceleration | 20Hz | Flight Mode | On change | Incoming Calls | On change |
| Gyroscope | 20Hz | Audio Mode | On change | Outgoing Calls | On change |
| Gravity | 20Hz | Battery Charge | On change | Incoming Sms | On change |
| Rotation Vector | 20Hz | Battery Level | On change | Outgoing Sms | On change |
| Magnetic Field | 20Hz | Doze Modality | On change | Notifications | On change |
| Orientation | 20Hz | Headset plugged in | On change | Bluetooth Device Available | Once every minute |
| Temperature | 20Hz | Music Playback | On change | Bluetooth Device Available ( Low Energy ) | Once every minute |
| Atmospheric Pressure | 20Hz | WIFI Networks Available | Once every minute | Running Application | Once every 5 seconds |
| Humidity | 20Hz | WIFI Network Connected to | On change | Location | Once every minute |

On a technical level, all data are generated as time-series, consisting of a tuple composed of a timestamp and one or more values. As briefly mentioned above, the smartphone generates and stores data locally before synchronizing it with the backend server for permanent storage. The device stores time-series tuples in a buffer in memory and as soon as the buffer is full, it is unloaded in a compressed and encrypted file on the device local storage, inside the application sandbox that prevents other applications from assessing them. Upon receiving the logs, the backend processes and stores them in a distributed

Cassandra database. On average, we expect a modern smartphone to generate 500MB per day of uncompressed data. A data collection with 500 participants would generate around 7.5TB, without redundancies and backups, in one month.

Once the data is stored, it is immediately available for analysis. The main way to access these data is to read them directly from the Cassandra database that due to its distribution and scalability, allows to reply to queries in a linear amount of time even with huge amounts of data (in the order of Tera Bytes). To make the data available to a broader audience, not only limited to computer scientists, pipelines were created to export the data, making use of Apache Spark, a distributed computation tool that reads the data directly from Apache Cassandra and writes them in files on a file system according to the Apache Parquet format. For the European Big Data Hackathon 2019 the participants were provided with an environment with a big data cluster, where they could use a distributed computing infrastructure powered by Apache Spark that naturally integrates with the Apache Parquet file format.

**5. Pilot Studies**

A pilot study is composed of six steps that the participants are asked to perform. These steps are:

1. Once the subject decides to participate, he is invited to fill out a personal questionnaire where the sociodemographic characteristics, the psycho-social information, together with their personal data (phone number, address, smartphone characteristics, etc.) are collected. While initially this was run separately, in the latest iterations of the studies this process was included directly in i-Log, without the necessity to use an external solution;
2. The subject is provided with a code (the same for every participant) so that to be able to initialize the i-Log application;
3. The subject is allowed to download the i-Log application from the Google Play Store (i-Log 2019) and install it on his personal Android smartphone (at the time this paper has been written, the iOS version was not yet available and currently under development);
4. The subject is required to insert the code that identifies the study into i-Log to start using it. Without this code, i-Log does not perform any operation. We decided to add this additional security layer in order to be sure about the participants of each study, forbidding external people to participate;
5. At the first execution, i-Log presents the user with an installation procedure (Figure 3). The objective of this procedure is to explain the purpose of the study, to formally ask the subject to read and to give consent to the privacy statement and to grant permissions to collect data from the personal device (both from a technical and a legal point of view).
6. Once the installation of i-Log is complete, the pilot study enters in its active stage. i-Log will now collect data from the smartphone's internal sensors and administer the Time Use Surveys (TUS). During this stage, the participants are asked to use the mobile application for a specified period of time (from days to months). During this period, a helpdesk is available for (technical) issues via email or phone, in case the participants encounter some issues with the application that cannot be solved by reading the provided written manuals.

HETUS uses a time resolution of 10 minutes for the recording of the time of the activities. However, in i-Log this value can be changed for every data collection campaign. Each set of questions is pushed to the application from a cloud server, to guarantee synchronization among the participants, and once received by i-Log it is shown as a notification. Even if the respondent is instructed to reply as soon as the notification is received, in a real-life study, this does not always possible. For this reason, the respondent can be given a limited or unlimited amount of time to reply to a time-episode once the notification is received. An important feature of the application is the possibility to monitor different aspects of the behaviour of the respondent concerning his answering behaviour to each question. In fact, it records the time elapsed between the time of the notification and when the subject begins to fill in the diary, and the time taken to complete the time-episode diary. This information is useful for testing the reliability of the respondent answers ([Bison et al. 2018](#)) and is an innovative aspect introduced by i-Log and not present in paper-based TUS.

During a study, the smartphone of the user is required to be online at times to receive the questionnaire initially or to synchronize the collected data with the cloud server. This is not a problem for modern smartphones since they are connected for most of the time to a network, either Wi-Fi or 3/4G. However, the application can work even if the connection is not available for long periods of time. If for example, the user is in a building without network coverage, the individual questions are not received when supposed to, but instead they are all delivered only once the device is back online. Each time-episode is composed of the same questions as the HETUS study (what are you doing? where are you? who are you with?) plus a fourth question that is a seven-point scale about the person's mood. For each question, the user is presented with a list of possible pre-coded activities, places and peoples. Additionally, for the activities, it does not collect secondary activities. These choices have been made to reduce the respondents' cognitive load and time necessary to reply. If people doubted whether their activity matched one of the predetermined categories, they could find additional explanation and examples in the user manual. In addition to the time use diary, the respondent was required to reply to two additional experience questions per day, one in the morning and one in the evening about their mood, with a seven-point scale.

## 6. The European Big Data Hackathon 2019

Between the 8th and 12th of March 2019, the second European Union (EU) Big Data Hackathon ([Eurostat 2019b](#)) took place alongside the 10th New Techniques and Technologies for Statistics (NTTS) conference in Brussels. Seventeen teams nominated by European National Statistical Institutes competed to develop a data analytics tool to address the annual challenge: *"How can innovative solutions for data collection reduce response* *burden and enrich or replace the statistical information/data provided by the time use survey?"*

The European Big Data Hackathon had three main objectives:
- to solve statistical problems by leveraging algorithms and available data, by engaging with developers and data scientists across Europe, giving them the possibility to work with relevant data sets in order to generate new ideas and potentially contrive novel algorithms;
- to produce innovative products, including visualisation tools, developing prototypes that official statistics will be able to integrate at European and national level;
- to promote partnerships with the research community and the private sector, by raising awareness about big data initiatives in Official Statistics in Europe.

### 6.1. Data sources

Given the focus of the Hackathon, the teams were suggested to use big data, either the datasets provided for the event and/or some acquired by themselves. The organizers decided to provide personal big data about individuals, collected from their smartphones. The members of the teams and additional volunteers collected data before the hackathon using two frameworks, i-Log and myBigO. Additionally, the teams had at their disposal traditional time diary data. Part of the diary data was collected on purpose for the Hackathon, via i-Log, and part was collected previously for other purposes, from the Modular Online Time Use Survey (MOTUS) and the HETUS.

i-Log was used to provide the main dataset for the Hackathon. The time diary user input and the sensors data specifications were adapted for the Hackathon (see section 6.2). i-Log provided the only dataset where sensors big data and time diary data referred to the same sample.

The second source of big data was the one provided by myBigO, a framework developed in the context of BigO, an International European research project to fight against obesity (Diou et al. 2018). Through the myBigO mobile application, activity data together with information about mood, and pictures of meal and food advertisement was collected from volunteers. It contained raw data from sensors, pre-processed data and self-reported data. Sensor data included geolocation data and recorded signals from accelerometer, barometer, light, proximity sensor, relative humidity sensor and thermometer. The processed data contained the recognised (i.e. predicted) physical activity (steps, walking, jogging, biking…), recognized visited points-of-interest (POI) and recognized transportation mode for trips between detected POIs (foot, bike, car, bus, train). The prediction models used were the ones trained in the BigO project. The self-reported data contained the pictures of meals and the mood. myBigO did not collect time diary data.

A first source of time diary data was MOTUS, which is an online time use survey administered via a website and a mobile application. The dataset from MOTUS and used in the Hackathon was collected from a sample of teachers in primary and secondary schools in Flanders during one week in 2018 ([Minnen et al. 2018](#)). Participants encoded their activities with reference to a pre-specified classification of 81 work related activities 21 activities related to personal and free time, together with the exact start and finish time. For each activity, participants registered where and with whom they were. In the case of a travelling activity, they registered the mode of transportation used. In the case of a work-related activity, they registered which technical tools they were using, type of teaching platform and if they were satisfied with the working activity itself (scale 1 to 7). The dataset included individual validated data for 8.571 teachers.

A second source of time diary data was HETUS, which is a traditional paper-based time use survey ([Eurostat 2019a](#)). This dataset contained anonymised micro-data from HETUS wave 2010. HETUS wave 2010 consisted of 18 countries that had collected TUS data between 2008 and 2015 based on harmonised guidelines. From the 18 participating countries, 5 of them were included in the Hackathon datasets: Austria (AT, 8234 observations), Belgium (BE, 11118 observations); France (FR, 27903 observations); Hungary (HU, 8391 observations); and Norway (NO, 7882 observations). The data contained the background information of the individuals and their households, and a diary where every 10 minutes of the day the following information is recorded: main and secondary activity, where the activity took place, if the individual was alone or with someone and if ICT was used. Each data record (per diary day) contained a total of 1656 variables.

### 6.2. i-Log data collection

To optimise i-Log for the Hackathon, both the time diary input questions and the sensor data collection were adapted. The purpose of the adaptation was to allow the teams a wide range of possibilities for their analytic choices. One option was to allow the comparison of data collected by i-Log with data collected by the HETUS survey to some degree. The two surveys are very different, and a direct comparison is not possible; however, it offers insights on what is feasible with an innovative data collection such as smart surveys, versus a traditional data collection like the HETUS.

The app collected data through three modes: a one-time user input of personal background characteristics at the start of the collection phase, a regular user input and an automatic collection of sensor data throughout the whole data collection period (for privacy preservation and data protection see section 6.3).

Once, at registration time, the participants filled out personal background characteristics, namely gender, occupation, their main activity status, the employer, and the place of employment.

The regular user input was triggered once per hour. The choice of the frequency was driven by personal experience during an initial pilot phase, where a frequency of twice per hour was tested and considered too burdensome. Each hour the participants received a notification on their smartphone with four questions, and prompted to fill out information about:

- their activity "What are you doing?" with 19 answer categories such as sleeping, eating, working, etc.,
- the current location "Where are you?" with 13 categories such as home, workplace, restaurant, etc.,
- the mode of transport (if travelling with a selection of 8 categories such as car, bus, etc.),
- the persons being with the participants at the time of the question "Who is with you?" with 7 categories such as nobody, partner, friends, etc,
- and their mood "What is your mood?".

Each question included one open-ended category. If the participants did not fill out each user input, this created a backlog of questions that could be answered at a later stage.

The time diary input questions and its categories were adjusted as closely as possible to the HETUS survey questions, to allow maximum analytical possibilities for the Hackathon. It is evident that a different mode requires a revised design. When rewriting the questions to fit the screen of the smartphone, we shortened the questions and categories. Considering that some studies suggest that the quality of responses increase with the switch to app and online modes with respect to paper-based surveys ([Stella et al. 2018](#)), it can be assumed that this redesign reduces the response burden and improved the quality of the answers. The exact wording considered possible response burden and survey mode effects. Due to time restrictions, the user input questions could not be pretested as extensively as they could have been.

The possibility of the collection of the sensor data in i-Log is manifold. For the Hackathon the decision which sensor data to collect, took privacy preservation reasoning and collection needs into account. Automatically, the app collected the following sensor data: acceleration/ gyroscope/ gravity/ rotation vector/ magnetic field/ orientation/ temperature/ atmospheric pressure/ humidity/ proximity/ position/ Wi-Fi network connections/ running applications/ screen status, flight mode, battery status, doze modality/ headset, audio mode, music playback (no track info)/ notifications received, touch event/ cellular network info.

Before the actual data collection, the developed i-Log app for the Hackathon went through a small and brief experimental pre-test. This helped to improve the actual collection phase. Small initial communication problems that the volunteer participants in the experimental test encountered, like how to switch on permissions to receive the input questions were solved instantly without any disturbance for the data collection - thanks to the instant feedback from the backend system.

An important result of this small test showed that the user was prompted too frequently, to a point where the risk of dropouts was too high. Therefore, the frequency of user input was reduced to an hourly intervention. The backlog of questions created when the participant did not fill out the user input or was offline, created some irritation and was reduced to eight times. This is justifiable as in most cases it can be assumed that the participant has long stretches of the same activity, such as work or sleep, where he/she should not be asked to fill out the same activity too often. Those decisions kept the balance between the data collection needs on the one hand, and the volunteer data collection of the target volunteers on the other hand.

The target-volunteering participants for the data collection was the hackathon participants themselves, as well as other volunteers recruited by Eurostat and the participants. The target group was only persons using Android phones, as i-Log was only available for this operating system at the time of the hackathon. Eurostat colleagues received an article via the intranet describing the project in a convincing way and referring to the privacy statement. The registration for i-Log consisted of downloading the app from google play store, installation of the app and entering a four-digit access code. The data collection period was from 28 January until 10 February 2019. In total, 95 persons registered for the participation in the i-Log data collection experiment for the Hackathon.

At the end of the data collection period, 66 participants registered around 190 000hours of sensor data (between all the sensors and all volunteers). Besides the 29 volunteers who did not register any data, some of them did not report data every day. The number of volunteers reporting data throughout the 14 days of data collection varied between 39 and 52, with a clear decreasing trend in time (Figure 5).

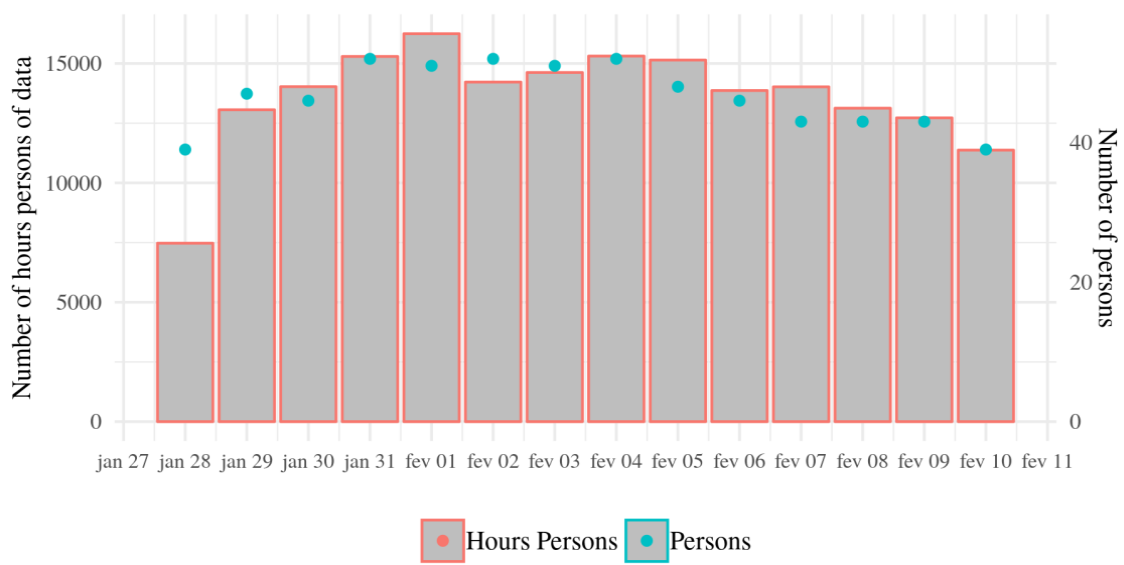

**Figure 5.** Amount of data collected (in hours and number of persons) per day

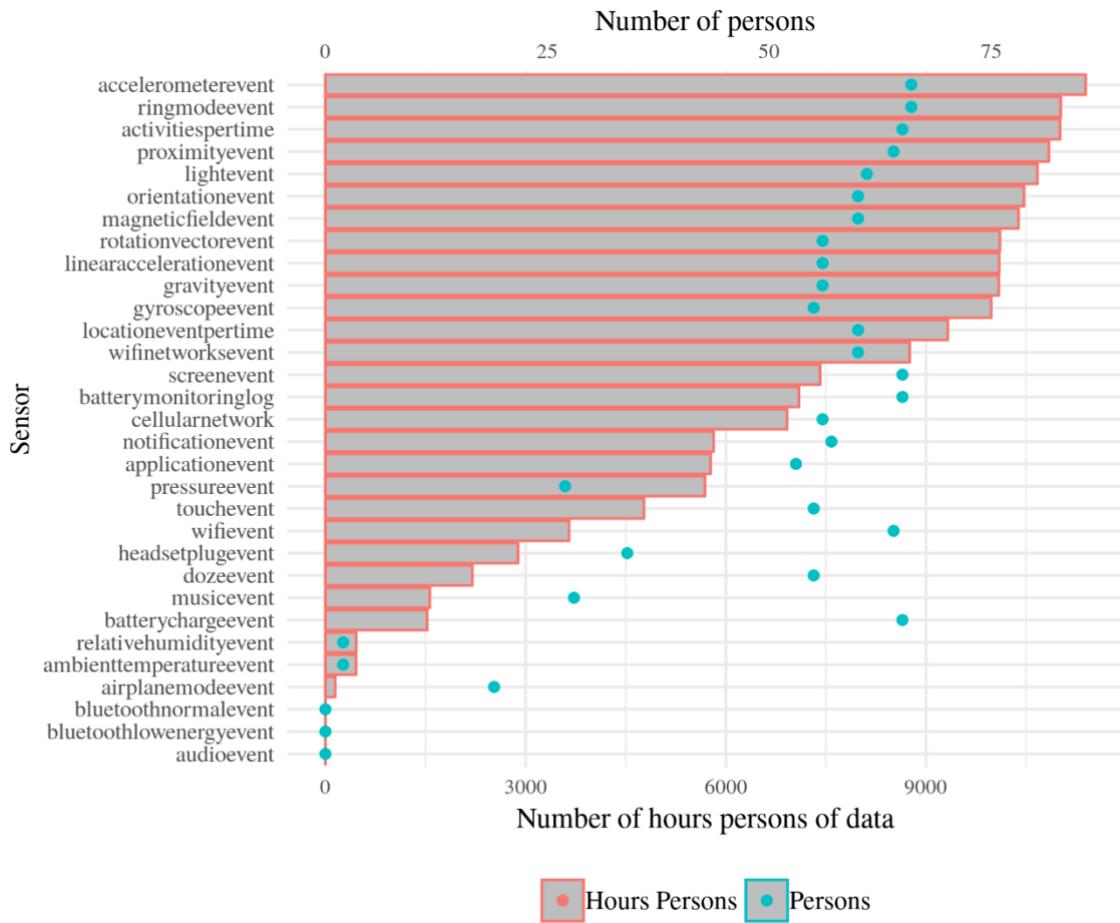

**Figure 4.** Amount of data collected (in hours and number of persons) from each sensor

Besides the data collected automatically by the application, each volunteer has registered on average around 15 diary hourly entries per day. In total, between all the volunteers 8548 entries have been registered.

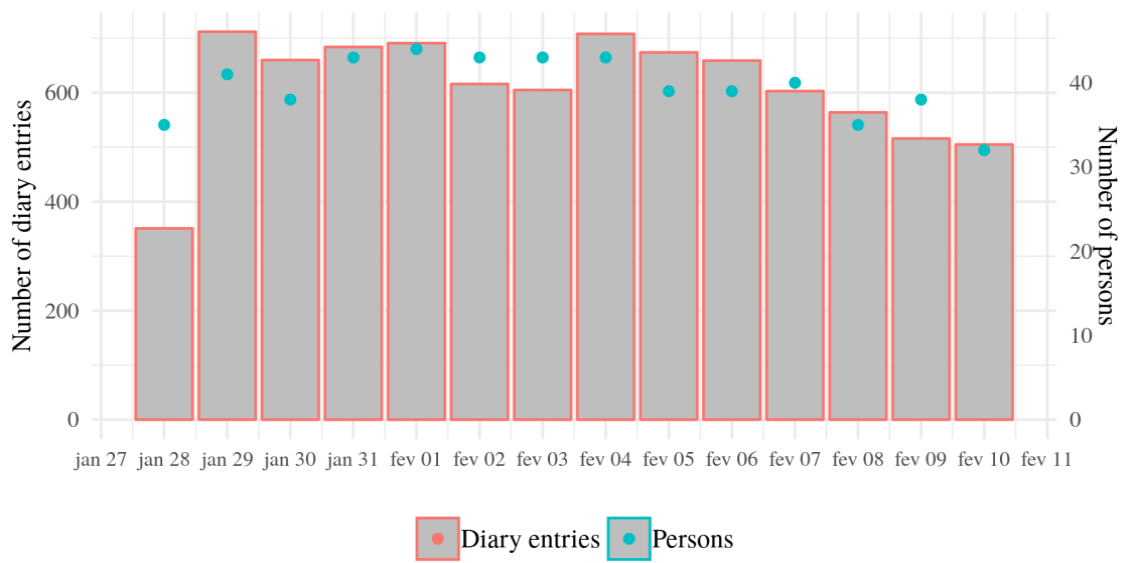

**Figure 6.** Amount of time diary entries collected per day

### 6.3. Privacy preservation and data protection

Personal big data poses particular challenges for the preservation of participants' privacy. In particular, geolocation data makes the re-identification of individuals in anonymised dataset relatively easy (De Montjoye et al. 2013). For this reason, privacy preservation and data protection placed particular importance in this experiment.

Privacy preservation is about the non-invasion of the private sphere of the data subject, i.e. the volunteer, meaning the non-disclosure of information he/she is not willing to share about him/herself. This was accomplished at three levels.

At the first level, this was done by allowing the volunteer to not share the information. In the case of the data actively inputted (i.e. the activities), the volunteer had the possibility of not answering. In the case of the data collected automatically (i.e. sensors) there were two mechanisms. The first one was by not giving permissions to the mobile application to access particular types of data (i.e. location data). This could be done in the Android settings and via the i-Log mobile application. The second mechanism was by giving the volunteer the possibility at any time to stop the collection of data (done via the mobile application).

At the second level, given that the volunteer chose to share his/her data, his/her privacy is preserved by minimising the risk that his/her identity is associated to the data, which was collected about him/her. His/her identity may be revealed by pieces of information which are public (or easy to obtain) and unique to her. In our case, this was mainly the email address obtained when the volunteer registered for the data collection. We minimised the risk by separating this identifying information from the data collected by the mobile application and by minimising the number of people and the cases where access was given to both types of data. The linking of both types of data needs to be possible, to comply with the legal obligation of giving the data subject the possibility to review, change and delete her data.

At the third level, given that even with only the data collected by the mobile application, it may be possible to indirectly identify the volunteer if additional information about her is obtained via other means, her privacy is preserved by minimizing the number of people who have access to the data and by restricting the time during which that access is granted. The people who had access to the data were the members of the 17 teams participating in the Hackathon, 2 persons in Eurostat (the data controller) and 2 persons responsible for the system administration of the backend. The access was granted until one month after the end of the Hackathon. On the 12 April 2019, all the data held by Eurostat was deleted and all individuals who had access to the data were legally obliged to delete any data still held (by the terms of use agreed by the participants in the Hackathon).

Data protection is about assuring that only the people who have been granted access to the data had access to the data, and not anyone else. This was done technically via the use of encryption during any data transmission and access control to any stored data. Besides those technical measures, this was done by imposing legal obligations to all individuals who had access to the data. The members of the teams participating in the Hackathon had to agree to the terms of use of the data during registration in the Hackathon, and these required them to "preserve the confidentiality of information". The staff of the App providers are bound legally by the contractual relationship between their organisations and the Commission and signed non-disclosure agreements. The staff of Eurostat is bound by the staff regulation of the European institutions.

### 6.4. Results

After the announcement of the challenge in the evening of the 8th of March 2019, the teams had two days to work on the development of a data product addressing the challenge. On the 11th of March 2019 in the morning, each team had 10 minutes to present their data product prototype.

The advantage of a hackathon is that is allows the generation of a relatively large number of ideas in a short time period. In total there were 17 teams participating in the Hackathon. Out of those, 6 teams used, or partly used, data collected by i-Log.

The Swedish team identified places of interest ("zones") and visualised trajectories in 3D, with the objective to improve the response burden for time use surveys by predicting the type of location (e.g. work, home). Based on i-Log data they created zones of mobility patterns over the day by using geo-positioning data. Every time of movement out of the defined zone triggers a question to the respondent to name their location. Over time this creates a multi-selection of activities and the system is able to make suggestions about which type of activities are being performed by the respondent based on sensor information.

The presented visualisation of trajectories can be used to reward and motivate participants. The tool can be used for machine learning to train the model before the user has to start filling out the questions in order to reduce the response burden and add to the transparency of the data collection process.

The Romanian team had the objective to improve the quality of time use survey by using smart survey data and analyse work-life balance variables based on TUS) data. The team explored all data available to assess if there are variables that could be removed to lower the response burden for the respondents. An imputation method for the transportation mode (on foot, public transport and by bicycle) was developed by using an R imputation package to visualise the variables and results in real-time. Further, an analysis of work-life balance variables was performed using HETUS and MOTUS data.

The Greek team explored how smart data can reduce the response burden of the time use survey by predicting the times of sleep activity by using data concerning the doze mode of the phone. Looking at i-Log data, the team identified peak hours were the doze mode was switched on peaks at 5 am and 3 pm were phones were at rest, and the doze mode was activated with a peak at midnight. Therefore, it was assumed that there is a correlation of doze mode and sleep. After cleaning some data (errors, short rest modes, etc.), a model was developed with i-Log data and then evaluated with the respondents' answers to check if it was indeed sleep. For future work, many other sensors of data are possible to perform similar models, for example on activities like cooking.

The Dutch team chose to focus on physical activities for time use surveys. As many other teams, they have identified the time diary as the main source of response burden and found that respondents tend to either decrease the amount of questions, or the time interval between the answers gets larger over time. To tackle this problem, the team proposed to label the activities automatically. After some exploration, the team found the i-Log data as the most promising to use. After some data cleaning, the team build a model for predicting activities form sensor data. Further, the team started to train the model and developed a convolutional neural network; however, some issues prevented further training. The labelling should have been more frequent, or closer to the activity, to allow an efficient search for the matching activity. Missing labels (non-response) in the accelerometer and the small sample size made it difficult to find the matching activity.

The Croatian team developed an app to visualise the relationship between HETUS data with i-Log data for the variables on activity and location. The objective was to compare sensor-based data information with traditional survey questionnaire information of the respondents. Two modules are possible to visualise in the app: activities and places of location. The team used open source software for their development, and the major difficulty the team faced was to standardise both sources in a way that it was usable for the app.

The Latvian team developed an app to compare data from i-Log sensor- and user input data on questions of activity and location. The idea was to find out if no significant movement of location might correspond to an inactive lifestyle, and if significant movement might indicate the mode of transportation. In sum, to predict activities from training using user input information. The team reported problems setting up the infrastructure, which led to insufficient time for the development in the Hackathon.

After the jury evaluated all proposals, they announced the six winners of the Hackathon. The six winners were: 1st Statistics Poland, 2nd Istat (Italy), 3rd ONS (UK), 4th Statistics Estonia, 5th Destatis (Germany), 6th Statistics Netherlands. At the award ceremony as part of the conference of New Techniques and Technologies in Statistics (NTTS 2019) the first three winners received their prize and gave the large audience a laureate lecture of their work;
- Statistics Poland received the 1st price for the creation of an open source prototype delivering a dashboard for the data analysis of the population time use.
- Istat from Italy received the 2nd price for the creation of "SMUTIS", an integrated open source environment for data analytics, visualisation and food classification.
- The 3rd price was given to the ONS, UK, for their development of a system to enrich the data collected via traditional questionnaire based surveys with an automatic processing of photos of meals taken by respondents.

The outcomes of the Hackathon and the event itself was a big success. Some teams are now in contact with HETUS and HBS production domains for further development and/ or integration of their prototypes.

To conclude, the use of sensor data to predict the location and to pre-fill the questionnaire would reduce the response burden immensely. Further, the time use survey can be enriched by visualisations, not only to increase the motivation for respondents to fill out the sometimes lengthy questionnaires (time diaries), but also to make the collected data more accessible to a wider audience, and promote the richness of information collected by the surveys.

**7. Lesson learned**

The pilot study presented in this paper for the European Big Data Hackathon 2019 is only one example of the kind of study that can be performed with the i-Log system. Other previous studies (Zeni et al. 2019), also as part of European H2020 projects (Maddalena et al. 2019), as well as planned studies in different countries around the world for 2019 and 2020 prove the feasibility of using i-Log in the field. Each study allows i-Log to improve the methodology and the system, as well as to introduce new functionalities.

Smartphones and related technologies are creating new opportunities and at the same time new challenges for TUS. They create new ways of sampling and recording information, which combines automatic and continuous data collection with limited or no human intervention. This is more accurate and decreases the burden of manually filling a traditional diary. It reduces the cost of performing a study and potentially increases the number of participants. At the same time, such hybrid solutions present many new methodological and technical problems. Mainly challenges are the selection of a sampling mode and the penetration rate, but it also raises issues on the ethical and privacy side, as well as technical challenges such as the usability, the connectivity, the design, the layout of the app, and the battery life of the devices.

From a technical point of view, each study results in a vast amount of data together with feedback collected from the participants that can be used to improve the i-Log application and its usability. In general, we can distinguish between two main categories of elements that can be improved: the time use survey part and the big data collection part. About the former, the main elements that the users reported as possible points to be improved are about extending the current functionalities of the application while replying. Examples are the possibility to reply "Same as the previous one", have a mechanism to automatically reply when is not possible to do it, e.g., while sleeping, at the cinema, etc. and to define standard routines to avoid replying to all the three sub-questions. Connected to the answering behaviour, some users highlighted that they needed a larger testing phase to understand the question wording, sequence, etc., as well as the categories and for the technical handling/user-interface of the app. About the big data collection part instead, we learned that an interesting feature to be introduced is about collecting data also from wearable devices in addition to the smartphone. In fact, many users now have a smartwatch or a smartband connected via bluetooth, and the data generated from them can provide additional insights about their activities and could also help filling those gaps when the user is not using the smartphone. The Bluetooth could also be used to detect nearby devices and detect physical networks of people in the real world. This functionality was originally present in i-Log but temporarily removed due to the high battery usage, but insights show that it should be restored because it allows richer data to be collected, which is important for time use analysis.

Concerning the backend part of the system, we learned some precious lessons from the feedback received and from what we could observe. First, by moving to the whole system to the cloud, we could reduce and optimize the resources needed to run a pilot study (and consequently the cost). In fact, in the cloud, everything is on demand and the system can scale up linearly depending on the load, while with standard servers they have to be bought advance since the buying process is long and complicated from a bureaucratic point of view. Additionally, when the pilot study is finished, the resources can be released, and the costs are reduced to zero. The migration to a cloud infrastructure also helped in improving the deployment phase of the whole architecture. With the Big Data Hackathon use case, we were able to move to a one-click-deployment pattern, where all the components of the backend were deployed instantly with a single user operation: this increases the reusability of our approach and reduces the time needed to run a new pilot study in a different site.

An additional element that improved the data collection was the helpdesk towards the final users, granting different levels of assistance. Level 1 composed of an exhaustive FAQ guide available online, level 2 an email address where a dedicated person could answer, and finally, a level 3 email support where the requests not satisfied by level 1 or level 2 could be answered directly by the engineers who built the system. Additionally, the helpdesk and the different levels showed the necessity of having a dedicated role in a pilot study called Field Supervisor. The responsibilities of this person are to monitor the pilot study through a dedicated backend interface that leverages on insights generated by the collected data. This data driven approach to support the field study helps preventing possible unwanted situations that bring a user to request assistance to the help desk. One example is a specific user who does not senddata to the backend server. In such a case, the field supervisor can be notified by the system of such behaviour, be proactive, contact the user and ask very focused questions to better understand the problem, or trigger a specific functionality. For example enabling the synchronization of the files over the Wi-Fi network.

## 8. Conclusions

In this article, we have described the experience with a pilot study of a smart survey in the context of the European Big Data Hackathon 2019, a satellite event of the NTTS conference, organized by Eurostat. The main tool used for this pilot was i-Log, which uses the smartphones of a pre-selected sample of respondents and combines two data collection modes based on active input from the subjects together with data collected passively from sensors inside the smartphone.

The results of this study look promising. i-Log proved to be able to carry out a real smart survey that combines multiple data sources which is not simply an extension of computer-assisted interviewing (CAI). It is a very new type of data collection, with elements of CAI interacting with a new set of user expectations. The challenges of these novel tools are still new and partially unknown: among them, we can mention the validation and completeness of the data due to malfunctions in the automatic systems.

We are at the beginning of a long and challenging journey. There are many issues to be addressed, from both a technical and a methodological point of view, like the exploitation of the data, and above all the protection of the privacy of the respondents. Nevertheless, there are great and unimaginable new opportunities that this new data collection tools offer us. The important aspect is to get started.

**Bibliography**

Barney, D. 2004. The Network Society. Cambridge: Polity Press.

Bison, I., and A. Scalcon. 2018. "From 07.00 to 22.00: A Dual-Earner Couple's Typical Day in Italy: Old Questions and New Evidence from Social Sequence Analysis." Gilbert Ritschard, Matthias Studer (edited by), Sequence Analysis and Related Approaches.

Bison, I., Zeni M., Busso M., Bignotti E., Giunchiglia F., and Veltri G. 2018. "More Than Meets the Eyes: Complementing Surveys with Mobile Phone Digital Data Trail." ESRA BigSurv18 Conference. Not available.

Castells, M. 2000. The Rise of the Network Society. The Information Age: Economy, Society and Culture, Volume I (2nd ed.). Oxford: Wiley-Blackwell.

Chenu, A, and L. Lesnard. 2006. "Time Use Surveys: a Review of their Aims, Methods, and Results." Archives Européennes de Sociologie / European Journal of Sociology, Cambridge University Press (CUP), 47(3): 335-359.

De Montjoye, Y.A., C.A. Hidalgo, M. Verleysen, and V.D. Blondel. 2013. "Unique in the crowd: The privacy bounds of human mobility." Scientific Reports 3: 1376. DOI: https://doi.org/10.1038/srep01376

Diou, C., I. Ioakeimidis, E. Charmandari, P. Kassari, I. Lekka, M. Mars, C. Bergh, et al. 2018. "BigO: Big Data Against Childhood Obesity." European Society for Paediatric Endocrinology. Vol. 89.

Dumazedier, J. 1975. "The Use of Time. Daily activities of urban and suburban population in twelve countries. Edited by Alexander Szalai." Revue française de sociologie 16 no. 1. 125-129.

European Statistical System Committee (ESSC). 2013. "Scheveningen Memorandum-Big Data and Official Statistics."

European Statistical System Committee (ESSC).. 2018. "Bucharest memorandum on Official Statistics in a datafied society (Trusted Smart Statistics)." DGINS Conference.

Eurostat (European Commission). 2009. "Harmonized European time use surveys, Guidelines 2008. Methodologies and Working Papers." Luxembourg: Office for Official Publications of the European Communities.

Eurostat (European Commission). 2019. European Big Data Hackathon. Available at: https://ec.europa.eu/eurostat/cros/system/files/european_big_data_hackathon_2019_-_description_20181119.pdf (accessed April 2020).

Eurostat (European Commission). 2019. "Harmonised European Time Use Surveys (HETUS) - 2018 Guidelines." Eurostat Manuals and Guidelines.

Fernee, H., N. Sonck, and A. Scherpenzeel. 2013. Data Collection with Smartphones: Experiences in a Time Use Survey. Tilburg University.

Fernee, H., and N. Sonck. 2014. "Measuring Smarter - Time-Use Data Collected by Smartphones." International Journal of Time Use Research 11(1): 94-111.

Gershuny, J. 2015. "Time Use Research Methods." International Encyclopedia of the Social and Behavioral Sciences (Second Edition) 24.

Gershuny J., Sullivan O.,. 2019. What We Really Do All Day: Insights from the Centre for Time Use Research. Penguin Books Ltd.

Gilbert G., and Becker G. S. 1975. The allocation of time and goods over the life cycle. NBER Books.

Giunchiglia, F., M. Zeni, E. Gobbi, E. Bignotti, and I. Bison. 2017. "Mobile Social Media and Academic Performance." The 9th International Conference on Social Informatics (SocInfo 2017), September, 2017. Oxford, UK. Available at: https://link.springer.com/chapter/10.1007/978-3-319-67256-4_1 (accessed April 2020). DOI: https://doi.org/10.1007/978-3-319-67256-4_1

Giunchiglia, F., Zeni, M., Gobbi, E., Bignotti, E. and Bison, I., 2018. "Mobile social media usage and academic performance."" Computers in Human Behavior 82: 177-185. DOI: https://doi.org/10.1016/j.chb.2017.12.041

Groves, R. M. 2011. "Three Eras of Survey Research." Public Opinion Quarterly 75(5): 861–71. DOI: http://doi.org/10.1093/poq/nfr057

Groves, Robert M., Floyd J. Fowler Jr, Mick P. Couper, James M. Lepkowski, Eleanor Singer, and Roger Tourangeau. 2011. Survey methodology. John Wiley and Sons.

GSM Association. 2018. The Mobile Economy Europe 2018. Available at: https://www.gsmaintelligence.com/research/?file=884c77f3bc0a405b2d5fd356689be340anddownload (accessed April 2020)

Hatuka, T., and E. Toch. 2017. "Being visible in public space: The normalisation of asymmetrical visibility." Urban Studies 54(4): 984-998. DOI: https://doi.dox.org/10.1177/0042098015624384

Hellgren, M. 2014. "Extracting More Knowledge from Time Diaries?" Social Indicators Research 119(3): 1517–1534. DOI: https://doi.org/10.1007/s11205-013-0558-6

i-Log. 2019. i-Log on the Google Play Store. Available at: https://play.google.com/store/apps/details?id=it.unitn.disi.witmee.sensorlog (accessed April 2020)

Juster, F., and Stafford, F. 1991. "The Allocation of Time: Empirical Findings, Behavioral Models, and Problems of Measurement." Journal of Economic Literature 29(2): 471-522.

Juster, F., Ono, H. and Stafford, F. 2004. Changing Times of American Youth: 1981-2003. Institute for Social Research. University of Michigan.

Kramarczyk, J. 2015. "Spending Time on Media - Results of Using Multitasking Frequency Questionnaire In Poland." International Journal of Time Use Research 12(1): 153-190.

Kramarczyk, J., and M. Osowiecka. 2014. "Time is Running Differently on the Internet." International Journal of Time Use Research 11(1): 94-111. DOI: http://doi.org//10.13085/eIJTUR.11.1.94-111

Link M.W., Murphy Joe, Schober Michael F., Buskirk Trent D., Childs Jennifer Hunter, Tesfaye Casey Langer. 2014. "Mobile Technologies for Conducting, Augmenting and Potentially Replacing Surveys: Executive Summary of the AAPOR Task Force on Emerging Technologies in Public Opinion Research." Public Opinion Quarterly 78: 779-87.

Link, M. 2018. "New data strategies: nonprobability sampling, mobile, big data." Quality Assurance in Education 26(2): 303-314. DOI: https://doi.org/10.1108/QAE-06-2017-0029

Maddalena, Eddy, Luis-Daniel Ibáñez, Elena Simperl, Richard Gomer, Mattia Zeni, Donglei Song, and Fausto Giunchiglia. 2019. "Hybrid Human Machine workflows for mobility management." In Companion Proceedings of The 2019 World Wide Web Conference (WWW '19). May, 2019. 102-109. San Francisco, CA, USA. Available at: https://dl.acm.org/doi/abs/10.1145/3308560.3317056 (accessed April 2020). DOI: https://doi.org/10.1145/3308560.3317056

Merz J. 2009. "Time use and time budgets: Improvements, future challenges and recommendations. Society for the Study of Economic Inequality ECINEQ 125.

Minnen, J., I. Glorieux, T.P. van Tienoven, S. Daniels, D. Weenas, J. Deyaert, S. Van den Bogaert, and S. Rymenants. 2014. "Modular Online Time Use Survey (MOTUS)- Translating an existing method in the 21 st century." Electronic International Journal of Time Use Research 11(1). DOI: https://doi.org/10.13085/eIJTUR.11.1.73-93

Minnen, J., J. Verbeylen, and I. Glorieux. 2018. "Onderzoek naar de tijdsbesteding van leraren in het basis- en secundair onderwijs. Deel 1: Algemeen. [Time allocation of teachers in the primary and secondary school. Part 1: General]." Vlaamse Overheid, Brussel: Vakgroep Sociologie, Onderzoeksgroep TOR 57 blz.

Robinson, J.P. 1999. "The Time-Diary Method: Structure and Uses. In Time Use Research in the Social Sciences." New York: Academic/Plenum Publishers.

Robinson, John P. 2002. "The time-diary method." Time use research in the social sciences: 47-89. DOI: https://doi.org/10.1007/0-306-47155-8_3

Runyan, J. D., T. A. Steenbergh, C. Bainbridge, D. A. Daugherty, L. Oke, and B. N. Fry. 2013. "A smartphone ecological momentary assessment/intervention "app" for collecting real-time data and promoting self-awareness." PLoS One 8(8). DOI: https://doi.org/10.1371/journal.pone.0071325

Sorokin, P. A., and Clarence Quinn Berger. 1939. "Time-budgets of human behavior". Vol. 2. Not Avail.

Stella C., Fisher K., Gilbert E., Calderwood L., Huskinson T., Cleary A., and Gershuny J. 2018. "Using new technologies for time diary data collection: Instrument design and data quality findings from a mixed-mode pilot survey." Social Indicators Research 137(1): 379-390. DOI: https://doi.org/10.1007/s11205-017-1569-5

Sugie, N. F. 2018. "Utilizing Smartphones to Study Disadvantaged and Hard-to-Reach Groups." Sociological Methods and Research 47(3): 458–491. DOI: https://doi.org/10.1177/0049124115626176

The Nielsen Company. 2018. The Nielsen Total Audience Report Q1 2018. Available at: https://www.nielsen.com/content/dam/corporate/us/en/reports-downloads/2018-reports/q1-2018-total-audience-report.pdf (accessed April 2020)

United Nations. 2010. "In-depth review on time-use surveys, Economic Commission for Europe." Conference of European Statisticians, Note by the German Federal Statistical Office. ECE/CES/2010/25. Paris, 2018. Available at: http://unstats.un.org/unsd/demographic/sconcerns/tuse/ accessed April 2020).

Vilhelmson, B., Elldér, E., and Thulin, E. 2018. "What did we do when the Internet wasn't around? Variation in free-time activities among three young-adult cohorts from 1990/1991, 2000/2001, and 2010/2011." New Media and Society 20(8): 2898–2916. DOI: https://doi.org/10.1177/1461444817737296

Wang, X.H., Zhang, D.Q., Gu, T. and Pung, H.K., 2004. "Ontology based context modeling and reasoning using OWL". In IEEE annual conference on pervasive computing and communications workshops, March, 2004. 18-22. Orlando, FL, USA. Available at: https://ieeexplore.ieee.org/abstract/document/1276898/ (accessed April 2020). DOI: https://doi.org/10.1109/PERCOMW.2004.1276898

Zeni, M., I. Zaihrayeu, and F. Giunchiglia. 2014. "Multi-device activity logging." ACM International Joint Conference on Pervasive and Ubiquitous Computing. September 13-17, 2014. 299-302. Seattle, WA, USA. Available at: https://dl.acm.org/doi/pdf/10.1145/2638728.2638756 (accessed April 2020). DOI: http://dx.doi.org/10.1145/2638728.2638756/

Zeni, M. 2017. Bridging Sensor Data Streams and Human Knowledge. Trento: University of Trento. Available at: http://eprints-phd.biblio.unitn.it/2724/ (accessed April 2020).

Zeni, M., Zhang, W., Bignotti, E., Passerini, A. and Giunchiglia, F., 2019. "Fixing mislabeling by human annotators leveraging conflict resolution and prior knowledge." Proceedings of the ACM on Interactive, Mobile, Wearable and Ubiquitous Technologies 3(1): 1-23. DOI: https://doi.org/10.1145/3314419